\documentclass[twocolumn,tighten,twocolappendix,runningheada,12pt]{aastex63}
\pdfoutput=1 
\usepackage{amsmath,amstext}
\usepackage[T1]{fontenc}
\usepackage{ae,aecompl}
\usepackage[utf8]{inputenc}
\usepackage{apjfonts} 

\usepackage[figure,figure*]{hypcap}
\usepackage{enumitem}
\usepackage{bm}
\usepackage{pifont}
\usepackage{slashed}
\input{hyperlink-year-only-natbib-patch}

\newcommand*{\https}[1]{\href{https://#1}{#1}}

\newcommand{\cmg}{\text{cm}^2\,\text{g}^{-1}}
\newcommand{\Mpch}{\,h^{-1}\text{Mpc}}

\renewcommand{\eqref}[1]{Eq.~(\ref{eq:#1})}

\newcommand{\figref}[1]{Fig.~\ref{fig:#1}}

\newcommand{\tabref}[1]{Tab.~\ref{tab:#1}}

\graphicspath{{./}{figures/}}

\begin{document}

\title[Constraints on SIDM from weak lensing]{Constraints on Dark Matter Self--Interactions from weak lensing of galaxies from the Dark Energy Survey around clusters from the Atacama Cosmology Telescope Survey}

\author{Susmita~Adhikari$^*$}
\email{*susmita@iiserpune.ac.in}
\affiliation{Indian Institute of Science Education and Research, Pune, Maharashtra, 411008, India}

\author{Arka~Banerjee}
\affiliation{Indian Institute of Science Education and Research, Pune, Maharashtra, 411008, India}

\author{Bhuvnesh~Jain}
\affiliation{Center for Particle Cosmology, Department of Physics and Astronomy, University of Pennsylvania, Philadelphia, PA 19104, USA}

\author{Tae-hyeon~Shin}

\affiliation{Department of Physics and Astronomy, Stony Brook University, Stony Brook, NY 11794, USA}

\author{Yi-Ming~Zhong$^*$}
\email{*yimzhong@cityu.edu.hk}
\affiliation{Department of Physics, City University of Hong Kong, Kowloon, Hong Kong SAR, China}
\affiliation{Kavli Institute for Cosmological Physics, University of Chicago, Chicago, IL 60637, USA}


\begin{abstract}
Self--interactions of dark matter particles impact the distribution of dark matter in halos. The exact nature of the self--interactions can lead to either expansion or collapse of the core within the halo lifetime, leaving distinctive signatures in the dark matter distributions not only at the halo center but throughout the virial region.
Optical galaxy surveys, which precisely measure the weak lensing of background galaxies by massive foreground clusters, allow us to directly measure the matter distribution within clusters and probe subtle effects of self--interacting dark matter (SIDM) throughout the halo's full radial range. We compare the weak--lensing measurements reported by \cite{DES:2021qzb}, which use  lens clusters identified by the Atacama Cosmology Telescope Survey and source galaxies from the Dark Energy Survey, with predictions from SIDM models having either elastic or dissipative self--interactions. To model the weak--lensing observables, we use cosmological N-body simulations for elastic self--interactions and semi-analytical fluid simulations for dissipative self--interactions. We find that current weak--lensing measurements already constrain the isotropic and elastic SIDM to a cross-section per mass of $\sigma/m<1~\cmg$ at a $95\%$ confidence level. The same measurements also impose novel constraints on the energy loss per unit mass for dissipative SIDM. Upcoming surveys are anticipated to  enhance the signal-to-noise of weak--lensing observables significantly making them effective tools for investigating the nature of dark matter, including self--interactions, through weak lensing. 
\\
\end{abstract}

\section{Introduction}
\label{sec:intro}

Dark Matter self--interactions were first proposed as a solution to the perceived small-scale problems of the cold collisionless dark matter (CDM) paradigm \citep{Spergel:1999mh, 2017ARA&A..55..343B}. While some of these issues have later been shown to have a possible resolution within the CDM paradigm with the inclusion of the effects of baryons and other systematic effects \cite[see e.g.,][]{Kim1812121}, the study of self--interactions of dark matter is broadly aimed at using the gravitational influence of dark matter on structure formation in the Universe to probe its particle properties that remain elusive to direct or indirect dark matter detection experiments \cite[see e.g.,][]{Buckley:2017ijx,2018PhR...730....1T, Drlica-Wagner190201055,Adhikari:2022sbh}.

Self--interacting dark matter (SIDM) naturally arises in particle physics, especially within so-called dark sectors, a collection of new particles that interact feebly with particles of the Standard Model. These new hidden particles can serve as the mediators for the non-gravitational self--interactions of dark matter. The simplest self--interactions are elastic and isotropic and naturally arise from a heavy mediator scenario once the mediator is integrated out. Other types of self--interactions might arise by introducing extra new light mediators. For instance, Yukawa self--interactions can be mediated by light dark scalars or dark photons, leading to non-trivial velocity and angular dependence~\citep{feng:2009mn,Ibe:2009mk,Tulin:2013teo}. 
The Yukawa self--interactions can be classified into the Born, classical, or resonant regime, depending on characteristics such as the coupling constant ($\alpha_\chi$), mediator mass ($m_\phi$), dark matter mass ($m_\chi$), and the relative velocity of incoming particles ($v_\text{rel}$). For the Born region ($\alpha_\chi m_\chi/m_\phi \ll 1$), the differential cross-section is:
 \begin{equation}
  \frac{\text{d}\sigma}{\text{d}\Omega} = \frac{\sigma_0}{\left(1+\frac{m_\chi^2}{m_\phi^2}\frac{v_\text{rel}^2}{c^2}\sin^2\frac{\theta}{2}\right)^2},\quad \sigma_0 = \frac{\hbar^2 \alpha_\chi^2 m_\chi^2}{c^2 m_\phi^4},
  \label{eq:longrange}
\end{equation}
where $\Omega$ is the scattering angle in the center-of-mass frame, $\hbar$ is the reduced Planck constant, and $c$ is the speed of light.
The interaction approaches a hard-sphere scattering at $v_\text{rel} \to 0$ and is Rutherford-like as $v_\text{rel}/c \gg m_\phi/m_\chi$. At other points in the parameter space, e.g., the resonant region with $\alpha_\chi m_\chi/m_\phi \gtrsim 1$ and $m_\chi v_\text{rel}/m_\phi c \lesssim 1$, Yukawa self--interactions exhibit more complicated velocity and angular dependence~\citep[see e.g.,][]{Gilman:2022ida}.

Sizable self--interactions of dark matter allow efficient energy redistribution within different regions of the dark matter halo. This redistribution could alter the halo's properties, such as  density and velocity dispersion profiles. The process starts with a net energy transfer from the halo's outskirts to the interior, forming a core at the center. The core keeps expanding until the inner halo becomes thermalized with the outskirts and a negative gradient of the velocity dispersion profile is built.  Afterward, as kinetic energy transfers out from the interior, the random motion of dark matter there can no longer support its gravity, and the halo begins to collapse, which in turn leads to more outward energy transfer and runaway collapse. This is the so-called core collapse or gravothermal collapse phase, culminating in a cuspy inner density profile and a higher velocity-dispersion for the inner halo.


The rate, and therefore, the time scales, for the core expansion and collapse depend on the type and strength of dark matter self--interactions. For elastic and isotropic self--interactions, one of the most robust constraints on the cross section strength comes from the observations of the Bullet cluster crossing, with $\lesssim 2\,\cmg$\citep{Randall:2008ppe} at $\nu_\text{rel} \sim 3000$ km$ $\,s$^{-1}$. For a cluster halo at median concentration, such a cross-section implies $\sim \mathcal{O}(100~\text{Gyr})$ for the halo to evolve into a deep core-collapse phase~\citep{Pollack:2014rja, Essig:2018pzq}. Therefore, under the assumption of elastic hard-sphere scattering, we expect clusters to be in the cored phase of their evolution. However, models such as dissipative SIDM introduce extra channels for more efficient energy dissipation, potentially leading to an accelerated core collapse that is cosmologically relevant~\citep{Essig:2018pzq,Huo:2019yhk}.  Such interactions are typical for SIDM mediated by light mediator particles~\citep{Finkbeiner:2007kk,Kaplan:2009de, Cyr-Racine:2012tfp,Cline:2013pca,Boddy:2014qxa,Foot:2014uba,Finkbeiner:2014sja,Boddy:2016bbu,Schutz:2014nka,Zhang:2016dck,Blennow:2016gde,Das:2017fyl}. For dissipative SIDM, the energy dissipation is characterized by three parameters: the elastic self--interaction strength, $\sigma/m$, the dissipative self--interaction strength, $\sigma'/m$, and the energy loss per dark matter mass, $E_\text{loss}/m$, due to the bulk cooling. Bulk cooling refers to the scenario where light dark particles (e.g. dark photons) that are generated during dissipative interactions escape the halo without re-absorption. The inelastic SIDM can be modeled similarly by setting the energy loss, $E_\text{loss}$, equal to twice the mass difference of the two near-degenerate dark matter states~\citep{Huo:2019yhk}. 

\begin{figure*}
    \centering
    \includegraphics[width=0.98\textwidth]{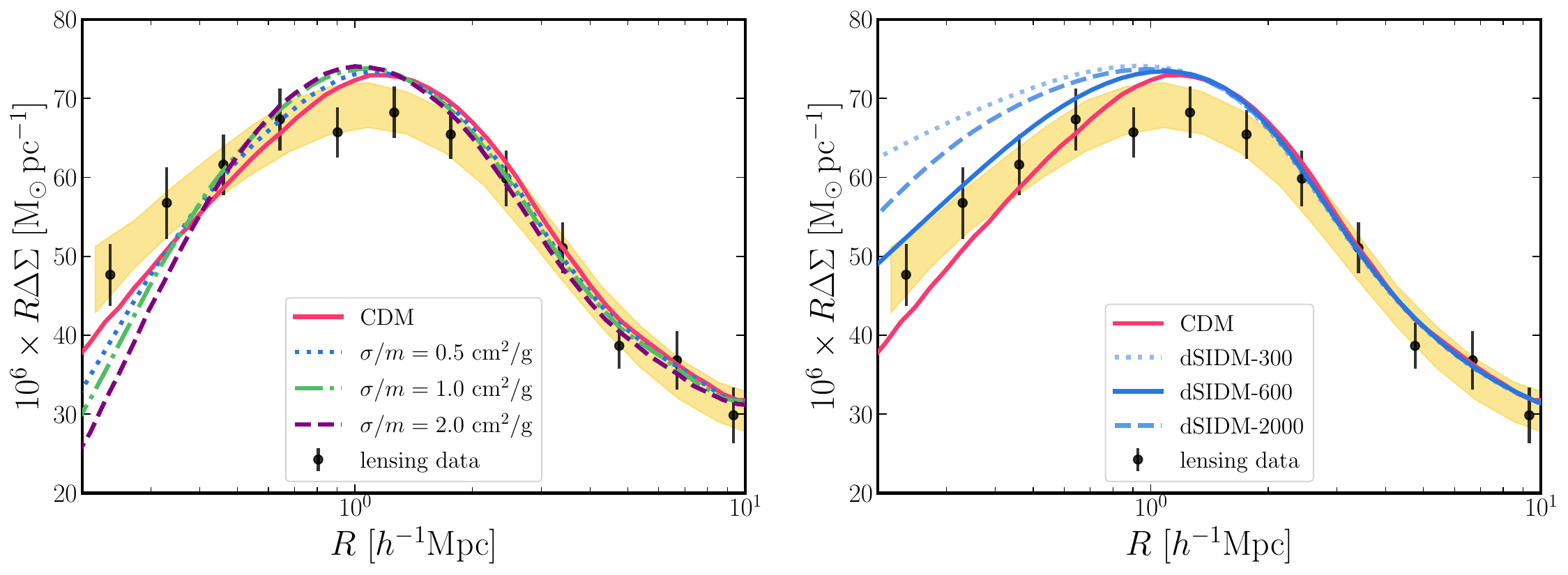}
    \caption{
     The stacked excess surface density profiles. The black dots show the lensing data, and the error bar indicates $1\sigma$ uncertainties. The orange-shaded regions show the $1\sigma$ model fit to the observational data. The left panel compares observational data with CDM and eSIDM with $\sigma/m=0.5, 1.0, 2.0\,\cmg$. The right panel shows the comparison with CDM and dSIDM with $\sigma/m=\sigma'/m=1\,\cmg$ and $\nu_\text{loss} = 300, 600, 2000\,\text{km}\,\text{s}^{-1}$.
      The density and projected radius are all in comoving units. See text for more details.
      }
    \label{fig:data_measurement}
\end{figure*}

This paper explores constraints on self--interacting dark matter with elastic and isotropic scatterings (eSIDM) and dissipative scatterings (dSIDM) using weak gravitational lensing observations of galaxy clusters. Galaxy clusters are some of the most exciting laboratories for studying the halo properties over an extensive range of scales (see \cite{Kravtsov:2012zs} for a review). The dark matter distribution of massive galaxy clusters that are of $\mathcal{O}(10^{14} M_\odot)$ are probed primarily through gravitational lensing and galaxy kinematics.   A range of studies have tried to infer the elastic hard--sphere scattering cross-section by measuring the deep inner core of a halo, primarily using strong lensing or stellar kinematics. Additionally, cluster scale probes include offsets of bright central galaxies in clusters \citep{Kim:2016ujt,DES:2023bzs} and offsets between the dark matter and stellar components of merging clusters\citep{Bradac:2006er, 2013ApJ...772..131D}. Some of these studies that use strong lensing find more stringent constraints on cross-section per mass of $\sigma/m<1\,\cmg$  \citep{Andrade:2020lqq, 2018ApJ...853..109E} at $v_{\rm rel}\simeq 1000 ~{\rm km} \, {\rm s}^{-1}$, while merging cluster studies put current constraints around the bullet cluster value at $2~\cmg$\citep{Wittman:2017gxn}.  Strong lensing probes the inner region of halos inside the typical Navarro–Frenk–White (NFW) scale radius~\citep{Gilman:2021sdr, Gilman:2022ida}, typically within $0.2 h^{-1}$Mpc.
In contrast, weak lensing studies of stacked clusters can be used to examine the outer regions of the halos. While the changes in the density profile are less pronounced in the outer regions than in the inner regions,  the halo properties are expected to be less influenced by baryonic effects and, hence, more robust. In addition, they can be effectively detected by current and forthcoming surveys, as we will elaborate in detail below.

Recently, \cite{DES:2021qzb} simultaneously measured the distribution of mass and the distribution of galaxies of massive galaxy clusters. The projected density profile of the lens clusters measured using weak lensing from this study spans a wide range of distances from the cluster center - from the inner scale radius of the halo at around $0.2 \Mpch$ to approximately  $20 \Mpch$. The profile  is a stacked measurement of the shear measured using galaxies from the Dark Energy Survey (DES) Y3 \citep{DES:2020aks} around a $\sim$1000 clusters identified using the Sunyaev--Zeldovich (SZ) effect from the Advanced Atacama Cosmology Telescope (AdvACT) Survey \citep{2021ApJS..253....3H}. Here, we use the profile from \cite{DES:2021qzb} (ranging from $0.2 \Mpch$ to $10 \Mpch$), to constrain eSIDM and dSIDM models. To confront the observational data, we leverage both cosmological N-body simulations as developed in \cite{Banerjee:2019bjp}\footnote{following the approach discussed in \cite{Rocha12083025}}, and semi-analytical fluid simulations as found in \cite{Balberg0110561, Koda11013097,Essig:2018pzq,Huo:2019yhk} to halo properties under a set of SIDM benchmarks. The latter method is necessary to simulate dSIDM halos as they may reach the deep core-collapse phase, a process that is computationally demanding for N-body simulations. For the sake of simplicity, both eSIDM and dSIDM are assumed to have velocity-independent scatterings. Conducting similar studies for velocity-dependent SIDM should be straightforward and will be performed in future studies.

The structure of our paper is as follows: Section \ref{sec:data} describes the lensing measurements of cluster profiles. Section \ref{sec:simulation} describes the simulations of SIDM halos for different SIDM models. Section \ref{sec:profiles} shows the resulting halo profiles for a set of SIDM benchmarks. Section \ref{sec:constraints} presents our constraints for different SIDM models and projects sensitivities of future surveys. We conclude in Section \ref{sec:conclusion}.

\section{Data and Weak lensing signal}
\label{sec:data}

This work uses the cluster weak--lensing measurements from \cite{DES:2021qzb}. In particular, we use the cluster--shear cross-correlation measurements to estimate the weak lensing of background galaxies by massive galaxy clusters detected by the SZ effect from the AdvACT survey, specifically the cluster catalog that was made publicly available as part of the fifth data release from the ACT \citep[ACT DR5, ][]{2021ApJS..253....3H}. 

The galaxy shape catalog has been measured using galaxies observed in the Dark Energy Survey (DES) \citep{2005astro.ph.10346T}.
In this work, we use data from the first three years of observation, particularly the DES Y3 gold catalog \citep{y3gold} similarly generated as in \citet{DrlicaWagner17}. The shapes of galaxies are measured using \textsc{Metacalibration} algorithm \citep{Huff2017, Sheldon2017}, where the gravitational shear, $\boldsymbol{\gamma}$, of galaxies is related to their ellipticity, $\textbf{\textrm{e}}$, by response, $\mathcal{R}$: 
\begin{equation}
    \langle \boldsymbol{\gamma} \rangle = \langle \mathcal{R} \rangle^{-1} \langle \textbf{\textrm{e}} \rangle,
\end{equation}
where the brackets represent the ensemble averages of the corresponding quantities. We refer readers to \cite{Y3shape} for details of the shear catalog. In addition, we use the corresponding photometric redshift catalog for our galaxy sample obtained by the Bayesian Photometric Redshifts (BPZ) algorithm \citep{bpz, Hoyle18}.
The details of the production of DES Y3 photo-$z$ catalog are described in \citep{y3gold}.

We make the same cluster selection as \cite{DES:2021qzb}. The cluster masses and redshifts are provided by the ACT catalog; we apply a cut on a signal-to-noise ratio (SNR), $\text{SNR}>4$, and redshift, $0.15 < z < 0.7$. This results in a total of 908 clusters in the $4552~{\rm deg}^2$ overlap between DES and ACT coverage. The mean cluster mass for this sample is  $3.1 \times 10^{14} \, h^{-1}M_{\odot}$. The full SNR and redshift distribution of the SZ clusters can be found in Figure 1 of \cite{DES:2021qzb}. 

\subsection{Measurement of weak lensing by clusters}
\label{sec:method}
The cluster--shear cross-correlation measurement is a direct estimate of the excess surface density, $\Delta\Sigma(R)$,\footnote{Here and henceforth, the distances and densities are \emph{comoving} quantities: $R$ is the comoving distance, $\rho(R)$ is the 3D comoving density, $\Sigma(R)$ is the projected comoving density, and $\Delta\Sigma(R)$ is the comoving excess surface density.} around the ACT cluster; this is defined as,
\begin{equation}
    \Delta \Sigma (R) = \frac{2}{R^2}\int_0^R \Sigma(R') R' \text{d}R' - \Sigma(R).
    \label{eq:delta_sigma}
\end{equation}
$\Sigma(R)$ is the projected mass density at a distance $R$ in the sky from the cluster center, which is related to the 3D mass density, $\rho(R)$, by
\begin{equation}
    {\Sigma} (R) = 2 \int^\infty_R  \frac{{\rho}(R') R'}{\sqrt{R'^2 - R^2}} \text{d} R'
    \label{eq:Sigma}
\end{equation}

The excess surface density is related to the measured tangential shear estimated from the ellipticity measurements of observed galaxies,
\begin{equation}
    \Delta\Sigma (R) = \overline{\gamma_t}(R) \, \Sigma_{\rm crit} (z_{\rm l},z_{\rm s}),
    \label{eq:Deltasigma}
\end{equation}
where $\overline{\gamma_t}(R)$ is the mean tangential shear at radius $R$ and $\Sigma_{\rm crit}$ is the critical density given by,
\begin{equation}
    \Sigma^{-1}_{\rm crit} (z_{\rm l},z_{\rm s}) = \frac{4\pi G}{c^2} (1+ z_{\rm l}) \chi(z_{\rm l}) \left[ 1 - \frac{\chi(z_{\rm l})}{\chi(z_{\rm s})} \right],
\end{equation}
where $G$ is the gravitational constant, $\chi(z)$ is the comoving distance to the redshift $z$, $z_{\rm l}$ the redshift to the lens, and $z_{\rm s}$ the redshift to the source galaxy.
We measure the tangential shear in 15 logarithmically spaced radial bins around the cluster centers ranging from $0.2 \Mpch$ to $30 \Mpch$.  Below $0.2 \Mpch$ the crowding of galaxies near the cluster center makes it difficult to measure the weak lensing signal. Further details about the measurement are given in Appendix \ref{ref:appendix_data}. In \figref{data_measurement}, we plot the measured excess surface density, $\Delta \Sigma$ times the comoving projected radius $R$. The measurement and the error bars are shown by the black points. The yellow shaded region shows the constrained dark matter density profile modelled using \cite{Diemer:2014xya} (DK14), the spread is obtained by sampling the $1\sigma$ distribution of fitted parameters \cite{DES:2021qzb}. 

In the following sections, we discuss the theoretical predictions for the dark matter density distribution in different models of self--interacting dark matter. 

\begin{table}[]
    \centering
    \begin{tabular}{c c c c c}
    \hline
        Name & $\sigma/m$ [cm$^2$/g]& $\sigma'/m$ [cm$^2$/g]& $\nu_\text{loss}$ [km/s] & Mtd.\\
        \hline
        CDM & -- & -- & -- & N\\
        \hline
        eSIDM & 0.2 & -- & -- & N\\
            & 0.5 & -- & -- & N\\
             & 1.0 & -- & -- & N\\
             & 2.0 & -- & -- & N\\
        \hline
        dSIDM-300 & 1.0 & 1.0 & 300 & F\\
        dSIDM-600 &  1.0 & 1.0 & 600 & F\\
        dSIDM-2000 &  1.0 & 1.0 & 2000 & F\\
        \hline
    \end{tabular}
    \caption{The benchmarks of eSIDM and dSIDM studied. We show the elastic and dissipative self--interaction strength in the second and third columns, respectively. The fourth column shows the velocity loss $\nu_\text{loss} \equiv \sqrt{E_\text{loss}/m}$ for dSIDM. The last column shows the simulation method used for each benchmark, cosmological DMO N-body simulation (N), and semi-analytical DMO fluid simulation (F).}
    \label{tab:benchmark}
\end{table}

\begin{figure*}[t]
 \centering
    \includegraphics[width=\textwidth]{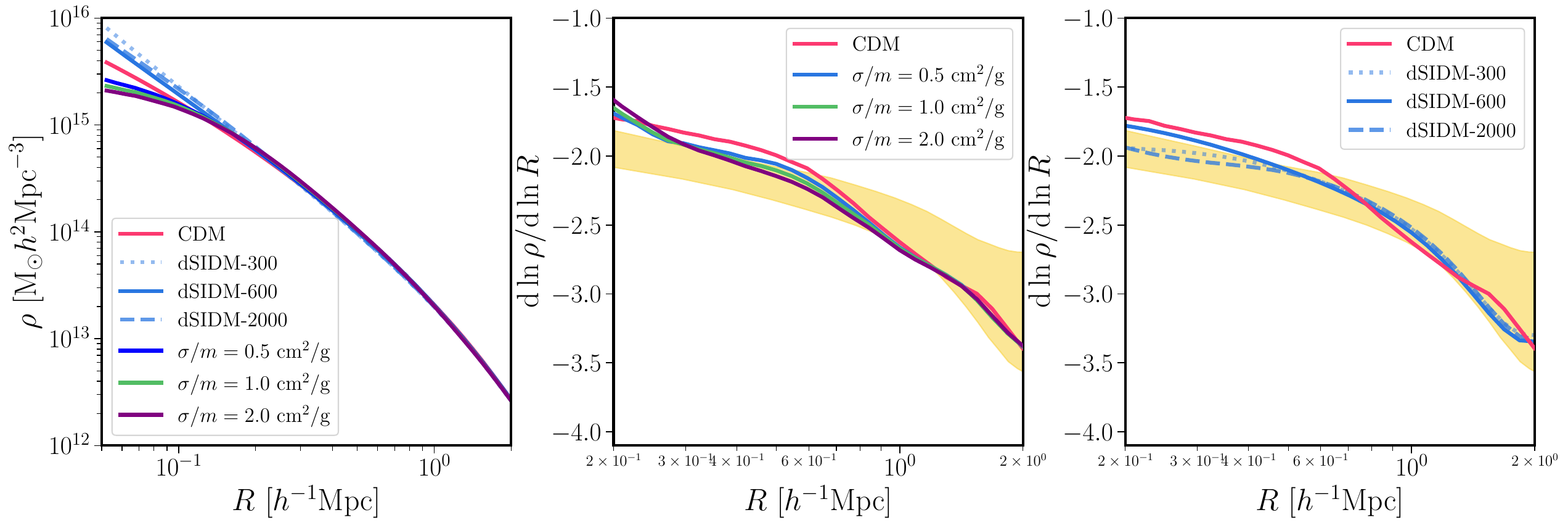}
\caption{The left panel shows the stacked 3D density profiles for all the benchmarks. The middle panel shows the resulting log-slope of density profiles for CDM and eSIDM benchmarks from N-body simulations. The right panel shows the results for dSIDM benchmarks from fluid simulations. The yellow shaded regions in the middle and right panels show the $1\sigma$ model fit to the observational data. The density and projected radius are in comoving units. 
}
\label{fig:3d_summary}
\end{figure*}

\begin{figure*}
    \centering
    \includegraphics[width=0.96\textwidth]{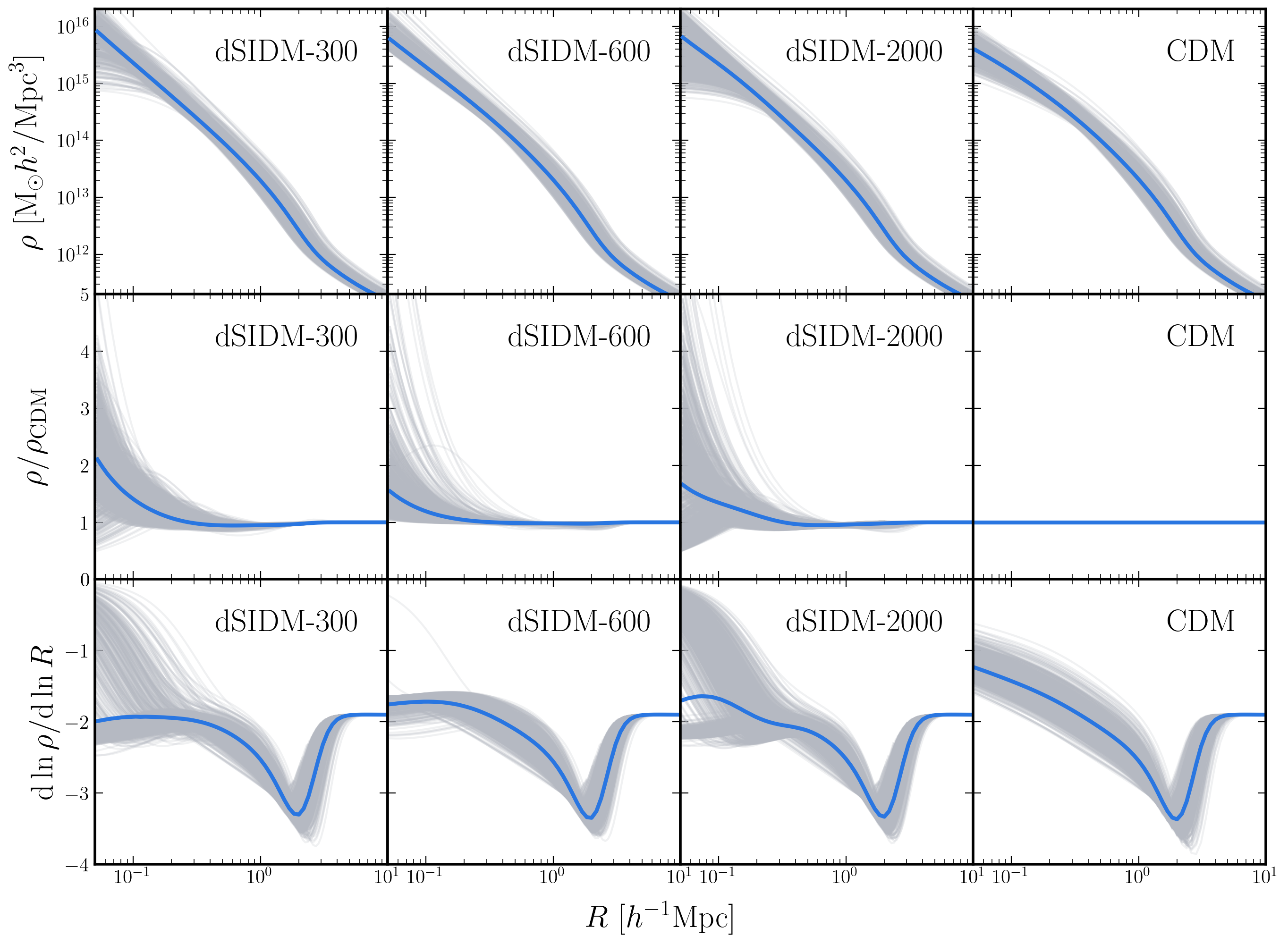}
    \caption{The first row shows the density profiles for each halo in the sample set (gray lines), along with the stacked density profiles (blue lines) under four different benchmarks: dSIDM-300 in the first column, dSIDM-600 in the second column, dSIDM-2000 in the third column, and CDM in the fourth column. The second row shows the relative density profiles to CDM, and the third row shows the log-slopes of the density profiles. The CDM profiles are obtained from the DK14 profiles generated by~\textsc{Colossus}
    }
   \label{fig:pre-stack}
\end{figure*}

\section{Simulations of self--interacting Dark Matter}
\label{sec:simulation}
We primarily employ two suites of SIDM simulations. For eSIDM, we conduct full cosmological dark-matter-only (DMO) N-body simulations based on \cite{Rocha12083025, Banerjee:2019bjp}.  Meanwhile, for the analysis of dSIDM (with dissipative scatterings on top of elastic and isotropic scatterings), we run fluid simulations for  DMO halos, following the recipe described in~\cite{Essig:2018pzq}. 

\subsection{N-body simulation}
\label{sec:nbody}
For the N-body simulations, we simulate a cosmological volume of $1000~h^{-1}$Mpc with $1024^3$ particles with the \emph{Planck} Cosmology ($h=0.68$, $\Omega_m = 0.31$, $\Omega_\Lambda = 0.69$ and $\rho_{\text{c},0} = 277.5 \, M_\odot h^2 \text{kpc}^{-3}$). This setup gives us a mass resolution of $7.78\times 10^{10} h^{-1} M_{\odot}$. The force softening length is adopted to be $0.015 h^{-1}$ Mpc. We include the non-gravitational self--interactions in an optimized version of \textsc{Gadget-2} based on \cite{Banerjee:2019bjp}. The interactions are modelled similar to \cite{Rocha12083025}, where two simulation particles ($i$ and $j$) that come within a distance of $h_{\rm SI}$ have a finite probability of interaction,
\begin{equation}
P_{ij}=\sigma v_{\rm rel}g_{ij}\Delta t
\end{equation}
where $v_{\rm rel}$ is the relative velocity between the two simulation particles, $\Delta t$ is a simulation time step, and and $g_{ij}$ is an overlap fraction defined as,
\begin{equation}
    g_{ij}=\int_0^{h_{\rm SI}}  W({\bf |x'|}, h_{\rm SI})W({\bf |x'+\delta x'_{ij}|}, h_{\rm SI})\,\text{d}^3{\rm x'}
\end{equation}
The weighted interaction kernel, $W$, is a standard kernel used in smooth particle hydrodynamics \citep{1985A&A...149..135M, Banerjee:2019bjp, Rocha12083025}. Apart from \citep{Banerjee:2019bjp} the simulations run with the above method were also used to study substructure lensing in \cite{Bhattacharyya:2021vyd} and to study Milky Way halos in \citep{Nadler:2020ulu} and \citep{Nadler:2021rpo}.

For this work we run simulations for CDM, and three eSIDM benchmarks with $\sigma/m = [0.2, 0.5,1,2]\, \cmg$, as listed in~\tabref{benchmark}. To get the halo catalogs, we run the \textsc{Rockstar} \citep{Rockstar} halo finder on our simulations. We extract cluster-mass halos from the simulation by matching the distribution of the observed sample; in particular, we choose a lower mass threshold such that the mean mass of the halo sample matches the observed value of $\overline{M_{500c}} =2.72\times 10^{14} h^{-1}$ Mpc. Where $M_{500c}$ refers to the mass within the radius that encloses $500$ times the critical density, $R_{500c}$. Note that all our halos have more that $1000$ particles within $R_{500c}$. We use halos from seven snapshots between $0.15 < z < 0.7$ and weigh them appropriately. We also construct a sample set of cluster-mass halos at the mean redshift of the observed clusters, $\overline{z}=0.48$, matching the observed distribution. The sample set from CDM will also serve as the cosmological initial condition for fluid simulation for dSIDM halos (Sec.~\ref{sec:fluid}).

\subsection{Semi-analytical fluid simulation}
\label{sec:fluid}
We employ semi-analytical fluid simulations to produce the modified halo density profiles for dissipative self--interactions. This fluid simulation method was originally developed for simulating globular clusters~\citep{1980MNRAS.191..483L} and later adapted for self--interacting dark matter halos~\citep{Balberg0110561}. The method obeys fundamental conservation laws, including the conservation of mass and energy, and makes common assumptions such as hydrostatic equilibrium and spherical symmetry for the dark matter halo.  After calibration with the results from N-body simulations, this approach can accurately depict the halo density profiles through the gravothermal evolution, encompassing both the core-expansion and the core-collapse phase~\citep{Koda11013097,Essig:2018pzq}. These calculations use far lower computational resources compared to N-body simulations, making it ideal for our task where we need to keep track of the evolution of $\sim 1000$ of halos from the cluster-mass halo sample set, taking into account their different initial density profiles, all the way to their deep core-collapse phases. 

We follow the set of equations as~\cite{Essig:2018pzq} for the halo evolution,
\begin{align}
    & \frac{\partial M}{\partial r} = 4\pi r^2 \rho, \frac{\partial (\rho \nu^2)}{\partial r} = -\frac{G M \rho}{r^2} \nonumber\\
&  \rho \nu^2  D_t \left(\ln \frac{\nu^3}{\rho}\right)=  -\frac{1}{4\pi r^2}\frac{\partial L}{\partial r} -C , \frac{L}{4\pi r^2 } = -\frac{\kappa}{k_B} \frac{\partial (m \nu^2)}{\partial r} \nonumber \\
 & C= \rho^2 \frac{\sigma'}{m} \frac{4 \nu \nu_\text{loss}^2}{\sqrt{\pi}} \left(1+\frac{\nu_\text{loss}^2}{\nu^2}\right)e^{-\frac{\nu_\text{loss}^2}{\nu^2}}, \kappa = \left(\kappa^{-1}_\text{lmfp} +\kappa^{-1}_\text{smfp}\right)^{-1} \nonumber\\
& \kappa_\text{lmfp} =\frac{3}{2\pi^{3/2}} \beta \frac{\rho \nu^3 \sigma k_B}{Gm^2}, \kappa_\text{smfp} = \frac{75 \pi^{1/2}}{64} \frac{\nu k_B}{\sigma},
\label{eq:fluid}
\end{align}
where $M$, $\rho$, $\nu$, $L$, and $C$ are enclosed mass, density, 1D velocity dispersion, luminosity, and the volumetric bulk cooling rate profiles, respectively. $k_B$ is the Boltzmann constant. $D_t$ denotes the Lagrangian time derivative. The reciprocal of the conductivity $\kappa$ is the sum of the reciprocals of the conductivities of the long-mean-free-path region $\kappa_\text{lmfp}$ and that of the short-mean-free-path region $\kappa_\text{smfp}$. Here, we assume both the elastic and dissipative self--interactions are hard-sphere interactions. $\beta$ is the calibration coefficient, which we set to $\beta\simeq 0.75$ based on calibrations with N-body simulations~\citep{Koda11013097}. The code is implemented in \textsc{C++}. In practice, the fluid equations are further processed to be equations with dimensionless variables. After obtaining dimensionless outputs, we map them to physical ones, assuming the same \emph{Planck} cosmology described in Sec.~\ref{sec:nbody}.

We conduct fluid simulations based on profiles obtained from realistic halos derived from the cosmological N-body simulations described in Sec.~\ref{sec:nbody}. We use the model profiles provided by \cite{Diemer:2014xya}, known henceforth as DK14 profiles, to get resolution over a large dynamic range. We use DK14 profiles for the mass and concentration distribution obtained from the 1148 halos in N-body simulations that match the observed mean mass and mean redshift of the observed sample as described in \ref{sec:nbody}. We generate these profiles using the \textsc{Colossus} package \citep{2018ApJS..239...35D}. In a nutshell, a DK14 profile includes two parts: (I) an inner term characterized by the Einasto profile, which is truncated near the virial radius to accommodate the splashback feature \citep{Diemer:2014xya, Adhikari:2014lna}, and (II) a smaller outer term formulated as a power-law profile to capture both the 2-halo or infall contribution and the constant background of cosmic mean dark matter density. For each selected halo in our simulations, we assume that its initial density profile, that is, the density profile at the time of the halo's formation, conforms to a DK14 profile. We set the power-law slope in the outer term to be $-1.9$, which appears to fit the stacked CDM profile well.  We then evolve each of these halos using the semi-analytical fluid simulation described above.

It is worth noting that the fluid simulation relies on the key assumption that the dark matter halo is in a (quasi-)hydrostatic equilibrium state at all times. For our setup, we assume that only the inner term of an initial DK14 profile participates in the gravothermal evolution, and the outer term remains static. This assumption holds validity as the inner term is virialized, and the outer term, corresponding to the infalling material, does not significantly contribute to the mass, particularly at a smaller radius. Based on this assumption, we only keep track of the gravothermal evolution of the inner term. We adopted the boundary conditions $M=0$ at $r=0$, and $M=M_\text{inner\, term}$ and $L=0$ at the outer boundaries for the fluid simulation. After obtaining the evolved inner term, we return the outer terms to obtain the evolved density profile. Note that the infalling dark matter, which contributes to the outer term, acts as a heating source for the SIDM halo and could potentially decelerate the gravothermal evolution if the accretion rate is high. We disregard this effect since we find the accretion rate of $10^{14}$ halos at $z=0.48$ is typically within  $\Gamma \equiv \text{d} \log M/\text{d} \log a  \sim 1.5-2$. Such rates are substantially less than $\Gamma \simeq 6$, the rate that would significantly slow down gravothermal evolution~(\cite{Ahn:2004xt}).

\begin{figure}
    \centering
    \includegraphics[width=1\columnwidth]{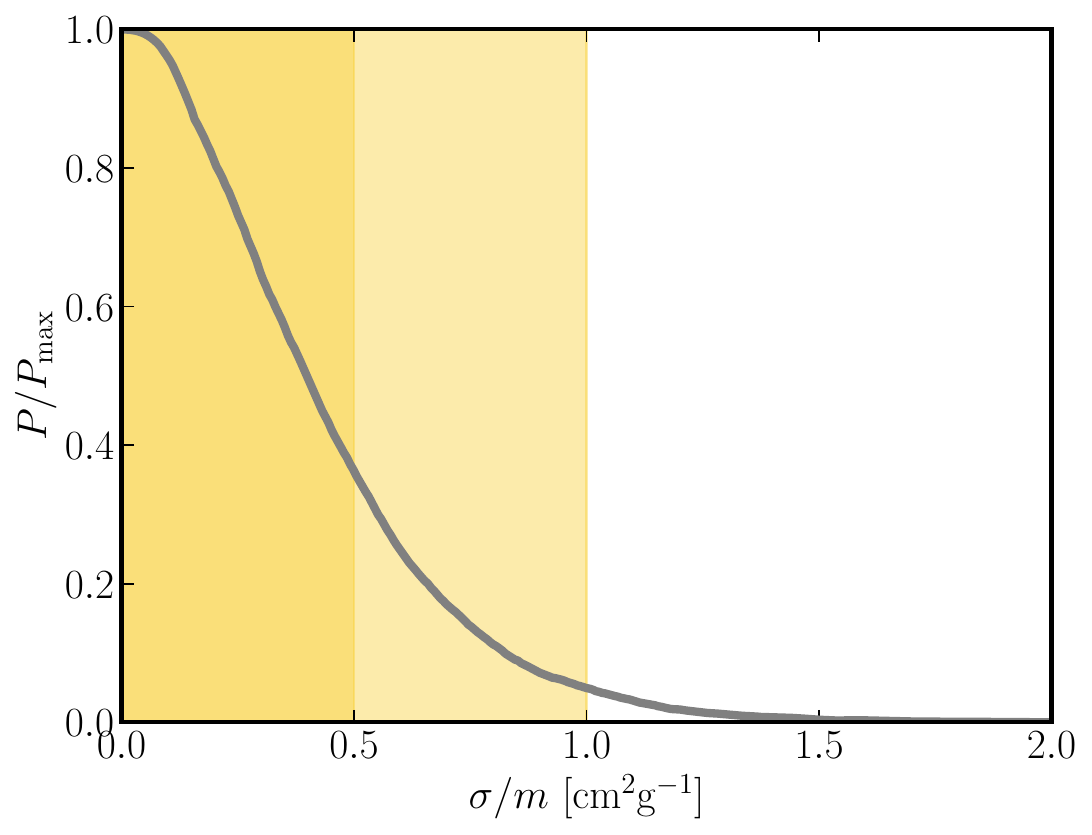}
    \caption{ Posteriors on the cross-section of isotropic, elastic scattering, $\sigma/m$ obtained using the weak lensing measurement of DES galaxies around 1000 SZ clusters selected from the ACT DR5 survey. The dark and light-shaded region shows the $67\%$ and $95\%$ confidence intervals.}
    \label{fig:current_constraint_plot}
\end{figure}

We consider four benchmarks for the fluid simulations, as listed in~\tabref{benchmark}, with the self--interacting cross sections saturating the Bullet cluster bound ($\sigma/m=\sigma'/m=1.0\,\cmg$). The three velocity loss for the dSIDM, $\nu_\text{loss}\equiv \sqrt{E_\text{loss}/m} = [300,600,2000]\,\text{km}\,\text{s}^{-1}$,  corresponds to an energy loss of $E_\text{loss} = [1, 4, 44.4]~\text{keV} \left({m_\chi}/{\text{GeV}}\right)$, respectively. For each benchmark, we run one simulation for each halo in the sample set. After obtaining the resulting halo density profiles, we compute the stacked density, the surface density, and the excess of the surface density profiles as described in Sec.~\ref{sec:method}.

\begin{figure*}
    \centering
    \includegraphics[width=0.48\textwidth]{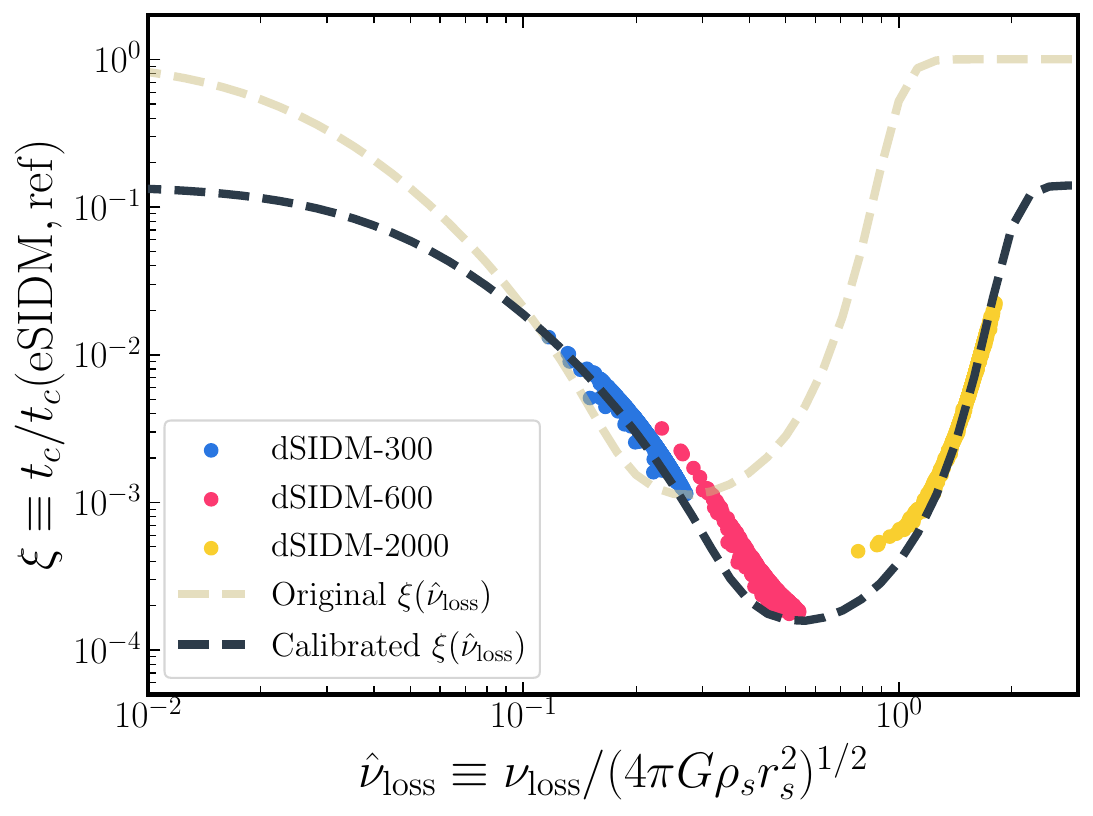}~\includegraphics[width=0.47\textwidth]{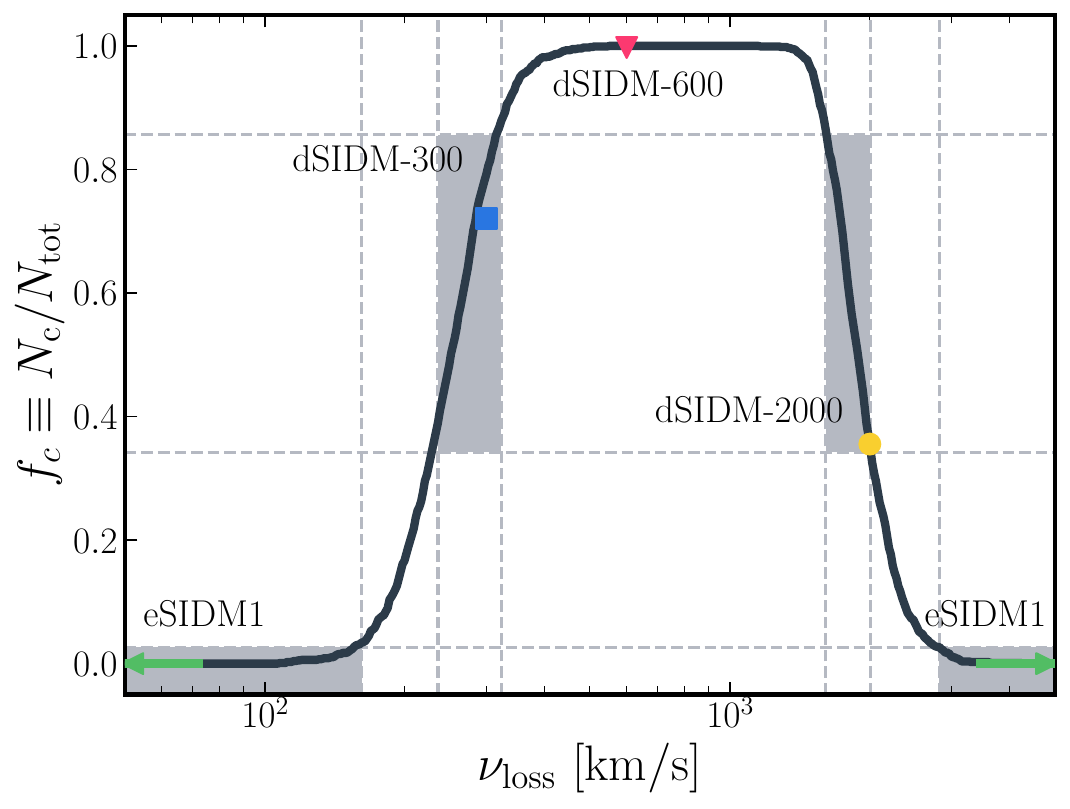}
    \caption{\emph{Left:} The simulated halos for the dSIDM-300/600/2000 benchmarks, in terms of $\hat \nu_\text{loss}$ and $\xi$, are shown as blue/red/yellow dots.
    The original $\hat \nu_\text{loss}$-$\xi$ relation from~\cite{Essig:2018pzq} is shown as the brown dashed line. The calibrated $\hat \nu_\text{loss}$-$\xi$ relation is shown as the black dashed line.   \emph{Right:} The fraction of collapsed halos, $f_c$, against $\nu_\text{loss}$ for dSIDM with $\sigma/m=\sigma'/m =1\,\text{cm}^2\text{g}^{-1}$. The projections of gray-shaded regions onto the $x$-axis indicate the parameter space where $\nu_\text{loss}$ is excluded with 95\% CL based on ACT$\times$DES weak lensing measurements. Blue/red/yellow dot shows $\nu_\text{loss}$ and $f_c$ for the dSIDM-300/600/2000 benchmark. Green arrows on left and right ends of the curve imply dSIDM approaches to eSIDM with $\sigma/m = 1\,\text{cm}^2\text{g}^{-1}$ (eSIDM1) at extremely small or large $\nu_\text{loss}$. See text for more details.}
    \label{fig:dSIDM_excl}
\end{figure*}

\begin{figure*}
    \centering
    \includegraphics[width=1.0\textwidth]{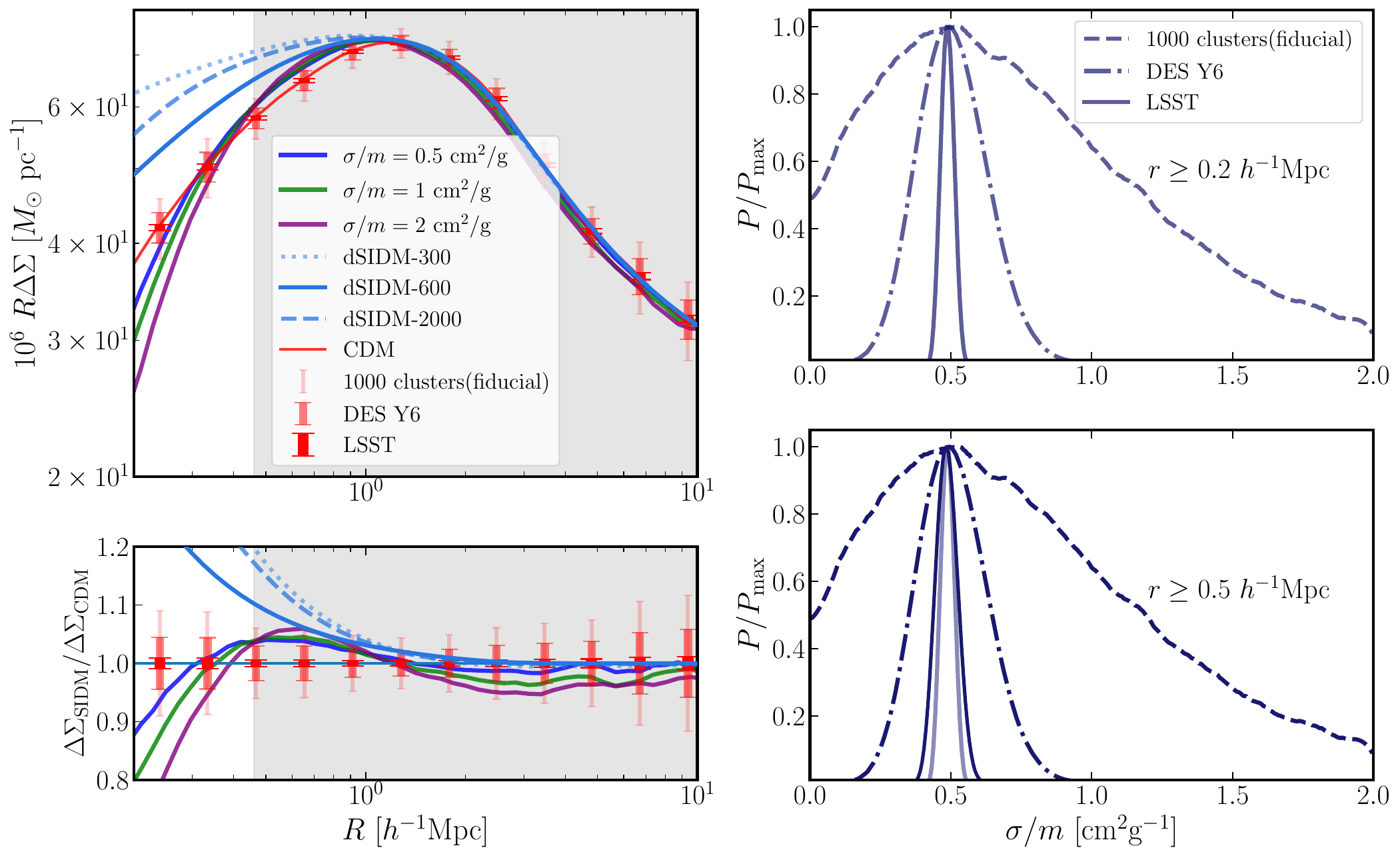}
    \caption{\textit{Left}: The top panel shows the measurement of $\Delta \Sigma (R)$ from lensing and how well we can estimate it using our fiducial sample of $1000$ clusters, DES Y6, and the full optical sample in an LSST-like survey. The bottom panel shows the measurements relative to CDM. \textit{Right}: Forecasts for constraints on eSIDM from future surveys for an injected value of $\sigma/m=0.5~ {\rm cm}^2 {\rm g}^{-1}$ using the full radial range of the signal above $0.2 h^{-1}$ Mpc. The dashed line at the top corresponds to the 1000 cluster sample used in this work. The dot--dashed and solid lines correspond to DES Y6 and LSST-like surveys, respectively. The bottom panel shows constraints if we only use the weak--lensing profile above $r\ge0.5~h^{-1}$ Mpc (shown by the grey shaded region in the left panel). The fainter solid magenta line is the same as the one in the top right panel for comparison.}
    \label{fig:future_constraint_plots}
\end{figure*}

In the next section we discuss the density profiles that arise in the different models of self--interactions and how they deviate from the standard cold dark matter scenario.

\section{Density Profiles of Dark Matter halos}
\label{sec:profiles}

In~\figref{3d_summary}, we show the stacked 3D density profiles, $\rho(r)$ and logarithmic slopes, $\frac{d\ln \rho}{d \ln r}$ for our halo sample. The profiles correspond to stacked quantities, where the halo mass distribution is matched to the observed distribution at the mean redshift of the observed sample. As expected, the density profiles for CDM are cuspy, approaching a slope of $-1$ near the center. For eSIDM N-body simulations, the halos develop a core near the center, and the inner densities get flattened within $\sim 0.3 h^{-1}$ Mpc. In contrast, the region adjacent to the core, at a slightly larger radius, becomes steeper than CDM as can be seen in the middle panel of~\figref{3d_summary}.  Note that if we allow arbitrary long evolution time for the eSIDM halo at these cross--sections, significantly past Hubble time, the eSIDM halos will evolve into the core-collapse phase; the region which is cuspier/steeper than the CDM density profile will be extend to smaller radii as the core collapses. 

For dissipative SIDM, the stacked profiles characteristically become steeper than CDM, as shown in the right panel of~\figref{3d_summary}. The reason is that, for each dSIDM benchmark, a substantial fraction of the halos quickly reach the deep core-collapse phase, which has a cuspier-than-CDM density profile. This is shown in the first three columns of~\figref{pre-stack} for the three dSIDM benchmarks. Each gray curve shows the density profiles (1st row), the relative density profiles to CDM (2nd row), and the log slopes of the density profiles (3rd row) for each of the 1148 cluster halos. The blue curves in each panel show the corresponding stacked profiles. In the last column, we show their CDM counterparts from the DK14 profiles generated by \textsc{Colossus}. The acceleration toward the core-collapse is the most significant for dSIDM with $\nu_\text{loss} \sim 600\,\text{km}/\text{s}$ (2nd column of~\figref{pre-stack}). Reducing the amount of dissipative energy loss permits fewer halos to collapse, as shown in the results for dSIDM-300 (1st column of~\figref{pre-stack}), where $v_\text{loss} = 300\,\text{km}/\text{s}$. Nevertheless, a significant portion of the sample set still reaches a collapse phase. Those halos have higher concentrations than cored ones. The results for dSIDM-2000 are similar to those of dSIDM-300. Not all halos in our sample collapse, but the underlying reason differs. For dSIDM-2000 (with $v_{\rm loss}= 2000 ~{\rm km/s}$), the energy loss is much larger than the typical kinetic energy of the dark matter particle. Therefore, it is difficult for the ground state particle to get excited to a higher energy state to release the light dark photons for energy dissipation. In other words, the dissipation process is Boltzmann suppressed, manifested as the $\exp(-\nu^2_\text{loss}/\nu^2)$ in the bulk cooling rate $C$. 

Besides the accelerating core collapse,  dSIDM benchmarks show a distinctly different inner halo density slope compared to eSIDM benchmarks due to the unique mechanisms of heat transport and energy distribution. In eSIDM, heat transport occurs through collisional conduction, depending on the velocity dispersion's radial gradient, which also affects the density profile. This results in a core-collapsed eSIDM halo with a typical inner density log-slope of about $-2.2$~\citep{1980MNRAS.191..483L}. In contrast, dSIDM features collisional conduction and bulk cooling for energy distribution during core collapse. The new channel reduces the reliance of the energy transport on the velocity dispersion gradient, leading to a less steep inner density profile in core-collapsed dSIDM halos, typically between slopes of $-2.2$ and $-1.5$. The exact slope depends on the interplay between bulk cooling and conduction, with stronger cooling associated with less cuspy slopes. Notably, these dSIDM halos still have cuspier-than-CDM inner density profiles. 

Focusing on specific dSIDM benchmarks, dSIDM-600 exhibits the strongest bulk cooling, resulting in all halos in the sample core collapsing. However, its net enhancement of density compared to CDM due to collapse is the least among the three benchmarks, as the inner slopes approach the values for the bulk-cooling limit, which are between $-1.5$ and $-2.0$ (see the second and third rows of~\figref{pre-stack}). On the other hand, dSIDM-300 has an overall weaker bulk-cooling effect, with some halos retaining their cored structures. However, in this case, the core-collapsed dSIDM-300 halos have steeper inner density profiles approaching $-2.2$, closer to completely core-collapsed eSIDM halos. dSIDM-2000, where cooling is Boltzmann suppressed, sees a larger fraction of halos remaining cored. The core-collapsed dSIDM-2000 halos have a steep inner slope comparable to the conduction cooling limit. Overall density enhancement in the case of dSIDM shows an M-shaped behavior as the cooling parameter, $\nu_{\rm loss}$, increases. As $\nu_{\rm loss}$ is increased from 0, the enhancement compared to CDM is maximized in dSIDM-300, then reduces in dSIDM-600, increases again in dSIDM-2000, and reduces again for  $\nu_{\rm loss}\gg 2000\,\text{km}\,\text{s}^{-1}$. Regarding $\Delta\Sigma$, a similar behavior is observed as shown in the right panel of Fig.~\ref{fig:data_measurement}.

Therefore, we note that the details of the shape of the stacked density profile and its log slope as a function of radius are sensitive to the comparative strength of bulk cooling vs. collisional conduction. Using the density information, we can constrain different SIDM models. 

\subsection{Comparison of simulations and data}

In this section we discuss the qualitative comparison between the data and simulations predictions. We obtain the dark matter density profile projected 2D space for each halo in our sample to construct $\Delta \Sigma$, the excess surface density observable. For the N-body simulation, we  project along the simulation $z$-axis.  The fluid simulation directly yields the 3D density profiles of the sample set. We average them to get a stacked 3D density, $\overline{\rho} (R)$, we further construct the stacked projected density, $\overline{\Sigma} (R)$ by integrating the 3D profile according to ~\eqref{Sigma} and the excess surface density, $\Delta \overline{\Sigma} (R)$, \eqref{delta_sigma}. 

The black points with error bars in Fig.~\ref{fig:data_measurement} shows the excess surface density, $\Delta \Sigma$, of massive galaxy clusters lenses that were found using the ACT survey as described in section \ref{sec:data}. The measurement of $\Delta \Sigma (R)$ is inferred from the tangential shear of the background galaxies from DES Y3. The yellow band shows the $1\sigma$ fits from the Markov chain Monte Carlo (MCMC) analysis when the data is fit to DK14 profiles as discussed in Section \ref{sec:data}. The different curves in the figure show a comparison of the theoretical models of dark matter to the data. The left panel shows comparisons with eSIDM benchmarks (based on N-body simulations), and the right panel shows comparisons with dSIDM benchmarks (based on the fluid simulations). 

In~\figref{data_measurement}, we note that the CDM profile differs from the measured signal in subtle ways. In particular, at small radii, $R<0.4\Mpch$, all the data points appear above the simulation prediction. \citet{DES:2021qzb} also find a similar discrepancy with mass curves derived from the Illustris simulations, which include baryons. Therefore, baryons do not explain the observed discrepancy. In the right panel, we note that adding in dissipative self--interactions increases the excess surface density at the measured scales, particularly between $0.2 \Mpch < R < 1.0 \Mpch$. The enhancement is due to the halos entering the collapse phase and the clustering of dark matter towards the center. Back to the left panel, we note that although the densities of eSIDMs are enhanced compared to the CDM in regions of $ 0.4 \Mpch < R < 1.0 \Mpch$ as dark matter is being pushed out due to core expansion, their inner regions within $R~=0.3 \Mpch$ are suppressed compared to the CDM, driving it further away from the observed data. The suppression of the density shows that the effects of the core are already manifesting themselves on these radii ($R \lesssim 0.3 \Mpch$), which is close to the scale--radius of the cluster mass halo. 

In the right two panels of \figref{3d_summary} we show the comparison of the logarithmic slopes of the different models to the inferred slope as a function of radius from data. The yellow band is from \citep{DES:2021qzb} and shows the 3D logarithmic slope of the DK14 profile that is fit to the lensing signal. The band corresponds to the $1-\sigma$ spread in parameter space. 

To summarize, the data shows a preference a weak-lensing strength larger than CDM in the inner three bins and a slope that is also steeper than CDM. The core-collapsed dSIDM halos have steeper inner profiles in this region inside the virial radius,  they also predict an overall larger weak--lensing signal compared to CDM. The completely core-collapsed benchmark (dSIDM-600) appears  closest to data. The eSIDM halos have steeper slopes than CDM at large radii but roll--over to a shallower core at small radii. In the next section, we test which models are currently consistent with data given the error bars and put constraints on the SIDM models. We use the dark matter profile from N--body or fluid simulations and the data points from  $\Delta \Sigma$ measurements shown in Figure \ref{fig:data_measurement}.

\section{Constraints on SIDM from current and future surveys}
\label{sec:constraints}

\subsection{Constraints on Elastic Self--Interactions}
 
 We run the eSIDM simulations at four cross section strengths between $0-2\,\cmg$ ($0~\cmg$ correpsonds to CDM) and evaluate the $\Delta \Sigma(R)$ profiles for each simulation, as discussed in the earlier sections. To get the model curves for other points in the $\sigma/m$ parameter space, we interpolate the simulation measurements of $\Delta \Sigma(R)$ as a function of $\sigma/m$ at every radial bin using a cubic spline. We use this to constrain the model with the ACT$\times$DES cluster weak lensing measurements. The results are shown in Fig. \ref{fig:current_constraint_plot}. We have used a basic Bayesian analysis assuming flat priors on $\sigma/m$ between $0-2\cmg$ to estimate the posteriors. We find that the current measurements from the ACT$\times$DES cluster weak lensing profile imply that $\sigma/m<1.05\, \cmg$ at $95\%$ confidence level (CL) and $\sigma/m<0.5\,\cmg$ at $67\%$ CL.
 The following section discusses our inferences about the dissipative dark matter models and the methodology used.

 Note that given the current error bars, we have not explicitly take into account the effect of cluster miscentering in this work. In principle a fraction of the observed halos can be miscentered affecting the density profile in the innermost bins, however we find in \cite{DES:2021qzb} the underlying DK14 profile marginalized over miscentering is well within the error bars of the lensing measurement. We check the effect of miscentering on the $\Delta \Sigma$ profiles from theory with values adopted from \cite{DES:2021qzb}; we assume $20\%$ of the clusters to be miscentered with a Gaussian distribution for miscentering with $\sigma=0.25$ Mpc $h^{-1}$, and find that the profile in the measured radial range is only marginally affected. In future work however, a jointly constraining these parameters will be important. The miscentering issue may also partly mitigated through using the outer regions, the implications of  which we will discuss in upcoming sections.
 
\subsection{Constraints on Dissipative Self--Interactions}

We evaluate the consistency of the dSIDM model with the ACT$\times$DES data. Intriguingly we find that dSIDM-600 benchmark demonstrates a better fit to the data compared to the CDM with a $\Delta \chi^2 = \chi^2(\text{dSIDM-600}) - \chi^2(\text{CDM}) = -4.75$. Both the CDM and the dSIDM-600 benchmark give good fits with, $\chi^2/{\rm d.o.f}$ of 0.9 and 0.5 respectively. In contrast, dSIDM-300 and dSIDM-2000 benchmarks showed poorer fits, with $\Delta \chi^2$ values wrt CDM of +19.73 and +5.54, respectively, leading to their exclusion at the 95\% CL.

Generally speaking, for a fixed $\sigma'/m$ and $\sigma/m$, the dependence of the stacked excess surface density on the velocity (energy) loss is intricate as described in Section \ref{sec:profiles}, the density enhancement is an M-shaped function of the energy loss parameter, $\nu_{\rm loss}$.
Our study of three dSIDM benchmarks reveals that these alone cannot interpolate the entire parameter space for a robust and exact constraint on the velocity (energy) loss of dSIDM. A more dedicated analysis of weak lensing constraints on dSIDM will be pursued in future studies. We have, however, used the fraction of collapsed halos to the total halo count, $f_c\equiv N_c /N_\text{tot}$, as an intermediate quantity to make a preliminary estimate of the current constraints on $\nu_\text{loss}$. The estimation is conducted in two steps: (1) we first estimate the fraction of collapsed halos $f_c$ for a given energy loss $\nu_\text{loss}$; (2) we estimate the stacked excess surface density $\overline{\Delta \Sigma}$ for a given $f_c$.
 
For step (1), we aim to establish a connection between $\nu_{\rm loss}$ and the collapse time, $t_c$, for all halos. We redefine $t_c$ for dSIDM in units of the collapse time in eSIDM, $t_c(\text{eSIDM,ref})$  as $t_c = \xi t_c(\text{eSIDM, ref})$, where $\xi$ represents the acceleration factor of dSIDM relative to eSIDM for the same $\sigma/m$. The reference eSIDM collapse time, $t_c (\text{eSIDM, ref}) = {150}/({\beta r_s \rho_s \sigma/m (4\pi G\rho_s)^{1/2}})
\label{eq:tcref}$, is given by~\cite{Essig:2018pzq} 
This approach is based on ~\cite{Essig:2018pzq}, who simulated dSIDM halos with the NFW initial profile and provide a detailed relationship between $\xi$ and $\hat \nu_\text{loss} \equiv \nu_\text{loss}/(4\pi G \rho_s r_s^2)^{1/2}$ (right panel of Fig.~2 in \cite{Essig:2018pzq}). 

To check this assumption, we plot the halos from dSIDM-300/600/2000 benchmarks on the $\hat \nu_\text{loss}$--$\xi$ plane. The distribution of these simulated halos collectively exhibits a ``U"-shaped relation, as shown in the left panel of~\figref{dSIDM_excl}. 
Next, we compare this data with the "U"-shaped curve from~\cite{Essig:2018pzq}, shown as the brown dashed line. Despite noticeable differences between our simulated halos and the original $\hat \nu_\text{loss}-\xi$ relation, possibly due to the difference in the initial profile modeling, the two can achieve a better agreement by simple rescaling the $\hat \nu_\text{loss}$ and $\xi$ variables of the original relation ($\hat \nu_\text{loss} \to 2 \hat \nu_\text{loss}$, $\xi \to 0.14 \xi$). The calibrated $\hat \nu_\text{loss}-\xi$ is shown as the black dashed line in the spanel. 

Once the acceleration factor is calibrated, we compute the collapse time for individual clusters in our sample and consequently the collapse fraction for the sample set. This gives us a relation between the collapsed fraction and the $\nu_{\rm loss}$ parameter through the acceleration factor. This is shown as the black curve in the right panel of~\figref{dSIDM_excl}. As expected, $f_c$ exhibits a bell-shaped dependency on $\nu_\text{loss}$.

For step (2), we calibrate $\overline{\Delta \Sigma}$ at each radial bin as a function of the on the fraction of collapsed halos, $f_c$. This is based on the distinctive halo profiles observed in dSIDM-300 and dSIDM-2000 benchmarks, as shown in the first and third columns of~\figref{pre-stack}. 
Each halo in a sample falls into two main categories: core-collapsed or cored. Despite significant differences in their density and log-slope profiles, the profiles within each category are largely consistent. Therefore, we propose that the combined $\overline{\Delta\Sigma}$ profile for dSIDM-300/2000 is effectively a mixture of $\overline{\Delta \Sigma}$ of cored eSIDM with $\sigma/m=1\,\cmg$ (eSIDM1) and that of completely core-collapsed halos. The mixing ratio is determined by $f_c$. We then obtain constraints on the parameter $f_c$ from the observed data.

Our analysis, which computes $\chi^2$ relative to CDM, finds that $f_c < 0.03$ or $0.34< f_c < 0.86$ are excluded with a 95\% CL. Those constraints translate into the limits of $\nu_\text{loss}$, using the $f_c$--$\nu_\text{loss}$ curve obtained in step (1). The resulting excluded parameter space is the $x$-axis projections of the shaded gray regions in the right panel of~\figref{dSIDM_excl}, corresponding to $\nu_\text{loss} < 1.6\times 10^2~\text{km}\,\text{s}^{-1}$,
$2.4\times 10^2~\text{km}\,\text{s}^{-1} < \nu_\text{loss} < 3.2\times 10^2~\text{km}\,\text{s}^{-1}$, $1.6\times 10^3~\text{km}\,\text{s}^{-1} < \nu_\text{loss} < 2.0\times 10^3~\text{km}\,\text{s}^{-1}$, and $\nu_\text{loss} > 2.8\times 10^3~\text{km}\,\text{s}^{-1}$. In the same panel, dSIDM-300/600/2000 benchmarks are represented as colored dots, and limits for eSIDM1 benchmarks are indicated with left and right arrows.

\subsection{Constraints from Future Surveys}

The current weak--lensing measurements are with an SZ cluster sample of $\sim 1000$ clusters. In the future, surveys like eROSITA \citep{Pillepich:2011zz}, LSST \citep{LSST09120201}, and Euclid \citep{Euclid} are going to increase the size of cluster samples to significantly large numbers. While eROSITA will give us X-ray clusters over the full sky, LSST and Euclid will primarily be finding optical clusters in the optical. The size of the optical cluster sample from LSST is likely to be $\sim 10^5$; such a sample will reduce the statistical errors by close to a factor of 10. In addition to increasing the cluster sample, the density of source galaxies for LSST will be much higher, nearly four times that of DES \citep{Chang:2013xja}, significantly reducing shape noise in the lensing measurements of galaxies. The relative amplitude of $\Delta \Sigma$ w.r.t. CDM and expected measurement errors in the future are shown in the left panel of \figref{future_constraint_plots}. Forecasts for future constraints are shown in the top right panel of Fig.~\ref{fig:future_constraint_plots} for DES year 6 (Y6) data and an LSST-like survey. To obtain the future forcasts we generate a mock lensing profile for eSIDM with $\sigma/m=0.5\, \cmg$, adding different levels of noise and use it to recover constraints on the cross-section. We find that one should be able to constrain the cross-section of elastic scattering to within $0.1 ~\cmg$  at $99\%$ CL using weak--lensing only. Similar improvements are expected in constraining the fraction of collapsed objects in the sample, giving comparable constraints on dissipative dark matter self--interactions.  

One of the caveats of the weak lensing measurement is the uncertainty in the estimated error bars of the boost factor (See Appendix \ref{ref:appendix_data} for discussion); this primarily affects the inner few points of the data deep inside the halo, where the differences between the SIDM signal and the CDM signal are maximized due to the onset of a core (in eSIDM) or cusp (dSIDM). Additionally, in the inner region, the light of the Brightest Cluster Galaxy (BCG) affects the lensing shape measurements out to at least $0.1 ~h^{-1}$ Mpc. As noted through this work, however, while there is a large signal in the inner region, differences exist throughout the multi-streaming or virial region of the cluster (or alternatively within the splashback radius). We utilize this fact and test how the constraints are affected by subsequently dropping measurements from the inner regions of the cluster. We report that in the near future, the outer profile can rapidly begin constraining the interaction cross-sections. In the bottom right panel of~\figref{future_constraint_plots}, for example,  we show the constraints we will get from using the dark matter weak lensing profile using data from  $R\ge0.5~h^{-1}$ Mpc. The dark blue solid line show the constraints obtained from an LSST--like survey when using only this outer region. Note that the LSST-like survey can be powerful enough to provide constraints that are nearly as good as those obtained from using the full profile down to $0.2 ~h^{-1}$ Mpc, implying that the outer profile alone will be a powerful probe of the nature of self--interactions. This mitigates not only issues discussed here, but also miscentering and baryonic effects that can become increasingly important at smaller radii. 

In this work we have primarily used the excess surface density $\Delta\Sigma$ to obtain constraints on the nature of dark matter. Observationally, the lens potential $\Delta\Sigma$ is derived by measuring of correlated distortions of background galaxies, characterized by the tangential shear,$\gamma_t$, around the clusters. The tangential shear is related to the underlying lens potential by the relation $\gamma_t=\Delta\Sigma/\Sigma_{\rm cr}$ as discussed in Section \ref{sec:data}. The denominator, $\Sigma_{\rm cr}$, is a normalizing factor that depends on the redshifts of the source and the lens galaxy samples. In the absence of information about the source redshift distribution, therefore, the lensing profile is simply uncertain by a normalization factor (we need the cluster redshifts, which are typically available, to obtain the distance in the lens plane). In such situations the shape of the weak lensing signal alone can be used to constrain the nature of dark matter. Figure \ref{fig:3d_summary}, which shows the slope as a function of radius, demonstrates this point clearly; The shapes of the lensing profiles are sensitive to the  cross-section of interaction. Having the amplitude, aditionally, allows us to break degeneracies but the shape alone can be used to rule out certain regions of parameter space. Surveys by the Euclid mission \citep{Euclid}, and others, may be able to provide cluster samples and galaxy shape measurements more easily and quickly than the photometric redshifts which require external multi-band photometry, allowing us to measure $\gamma_t$ making them powerful probes of dark matter self--interactions in the relatively nearer future. In a similar vein, the shape of galaxy number density as a function of radius can also potentially be used to constrain the shape of the dark matter distribution in clusters. \citep{DES:2021qzb} showed that the galaxy distribution follows the dark matter distribution closely barring a constant scaling factor related to galaxy bias.

\section{Conclusion}
\label{sec:conclusion}

In this study we have used the stacked weak lensing profile around massive galaxy clusters to probe self--interactions of dark matter. Using full N--body simulations for elastic, isotropic scattering (eSIDM) and semi--analytic fluid simulations for inelastic or dissipative scattering (dSIDM), we have shown how the complete profile shape within the virial region is sensitive to dark matter self--interactions. 

In the case of eSIDM, we observe that while the innermost region becomes cored, a steepening occurs at the adjacent region outside the core. This transformation affects the matter distribution out to scales of $\mathcal O(\text {1 Mpc}/h)$ and could serve as a distinguishing observational feature between different dark matter models. 

With regard to dSIDM, this is the first study that combines semi-analytical simulations with realistic initial conditions informed by CDM-simulations. The profiles of dark matter halos in dSIDM can become steeper than CDM within half the virial radius for our benchmark models and this occurs within a Hubble time due to an onset of core collapse. The steepness exhibits a complex dependence on the interplay between the bulk cooling and the conduction cooling process active in dSIDM halos, additionally the exact final profile depends on the dark matter halo properties, such as its concentration. 
The final \textit{stacked} density profile of dSIDM halos emerges from an interplay of two effects: the fraction of halos in the sample in the core-collapse phase vs. the exact value of the log-slope of the inner density profile of the core-collapsed halo. It is worth noting that the bulk cooling is most pronounced when $\nu_\text{loss}\equiv \sqrt{E_\text{loss}/m}$ approximates the virial velocity of the halo~\citep{Essig:2018pzq}.

We compare our results from simulations to state-of-the-art measurements from data. We use recent SZ--selected cluster data from ACT DR5, which uses DES Year 3 measurements of galaxy distortions to get the weak lensing signal and compute the excess surface density, $\Delta \Sigma$, around these clusters. We compare the measurements of $\Delta\Sigma$ to the lensing signals constructed from simulations at the mean redshift and mass of the cluster sample. We find that  at the smaller radii of the halo ($R<0.4\,\Mpch$), the dark matter density is cuspier than that observed in CDM--only simulations. Such a steepening in density profile is consistent with dSIDM models with strong energy dissipation (e.g., dSIDM-600). While eSIDM also predicts a steepening consistent with the measurements between $0.3 \Mpch < R < 1 \Mpch$, it quickly shows the signature of a core in the inner region which is deviates from with the data. 

Finally, we use the entire measured profile to obtain constraints on eSIDM. We forward model the lensing signal for the full redshift and mass distribution of the observed clusters using  cosmological N--body simulations. We obtain an upper limit of $\sigma/m< 1.05\,\cmg$ for elastic scattering at $95\%$ CL. This is comparable to the bullet cluster constraints\citep{Randall:2008ppe} and consistent with results from strong--lensing. We also place novel constraints on the energy loss per mass (or $\nu_\text{loss}$) for dSIDM with a fixed $\sigma/m$ and $\sigma'/m$ using the fraction of collapsed halo as intermediate quantity (\figref{dSIDM_excl}). We expect the constraints to be improved as we sample more benchmark points in future studies. Overall, we find that dSIDM-600 is the best fit to the data, with its $\chi^2/{\rm d.o.f}$ better than CDM. This model has an elastic and dissipative scattering cross-section of $1~\cmg$ and an energy loss corresponding to $E_\text{loss} = 4\, \text{keV} (m_\chi/\text{GeV})$ (Table \ref{tab:benchmark}). 

The improved precision of weak--lensing measurements around massive clusters from current and upcoming galaxy surveys like DES \citep{2005astro.ph.10346T}, LSST \citep{LSST09120201} and Euclid \citep{Euclid} will help put strong constraints on the dark matter interactions using weak--lensing only. LSST and Euclid will not only increase the cluster sample size, they will also significantly reduce shape noise errors due to their greater depth. Along with optical surveys, SZ-selected clusters from CMB experiments like Simons Observatory and CMB-S4 \citep{Abazajian:2019eic} and X-Ray telescopes like eROSITA \citep{Predehl:2010vx} will also provide much larger samples than the current ACT catalogs. The additional advantage of using such multi--wavelength information will be better constraints on systematics like cluster mass-observable relations, miscentering and projection effects. 

We note that our study does not take into account baryonic effects. Baryons, particularly through AGN feedback in clusters, can affect the dark matter density profile. However, the effects are not expected to be significant on the scales that we measure in the weak--lensing studies with the current level of measurement errors. This was also demonstrated in \citep{DES:2020aks, Dacunha:2021vdf} where the lensing and galaxy profiles are compared to clusters in Illustris simulations \citep{Nelson:2018uso}. However, in the future as measurement errors decrease,  it will be important to model or marginalize over baryonic effects to get robust constraints on the nature of interactions of dark matter. Several works like \citep{Robertson:2018anx} have explored the cluster regime using both baryons and non-gravitational self--interactions. Currently, they are zoom--in simulations of individual cluster mass halos, in the future, it will be important to extend such study to larger cosmological boxes. Since hydrodynamical  simulations are expensive and the feedback models have significant uncertainties it will also be interesting to explore semi--analytical approaches that parameterize the baryonic feedback and SIDM simultaneously \citep{Kelkar_inprep}. It will also be useful to explore models like \citep{Zhong:2023yzk}, which include a central baryonic potential in their fluid simulations in the cluster regime.

The details of the dark matter profiles of cluster halos encode important information about the history of  halo evolution on different scales and can encode subtle effects of the nature of dark matter self--interactions. Weak lensing studies are increasingly going to become a powerful probe of dark matter -- they probe the regions of dark matter halos beyond the core that are less affected by the massive central galaxy potential and baryonic effects. Current and future galaxy surveys can study large sample of clusters to put strong independent constraints on the nature of dark matter on the mass and velocity scale of clusters. In fact, the weak--lensing measurements of outer profiles of halo in general, extending down to groups and galaxy mass regimes \citep{Luo:2023tal, Thornton:2023gju}, will become powerful complementary probes of dark matter self--interactions allowing us to explore the velocity-dependence of dark matter self--interactions.

\section*{Author Contributions}

\textbf{S.A. $\&$ Y.Z.}: project conceptualization, formal analysis, methodology, software, investigation, validation and interpretation, writing- original draft, visualization. \textbf{T.H.S.:} data products from Dark Energy Survey and posteriors for weak--lensing measurements, interpretation, writing, editing. \textbf{A.B.:} simulations for analysis, conceptualization, interpretation.  \textbf{B.J.:} Conceptualization, interpretation and validation of results, editing.

\begin{acknowledgments}

 We thank Chihway Chang, Eric Baxter, and Stuart L. Shapiro for useful discussions. This research made use of computational resources at the University of Chicago’s Research Computing Center and the PARAM Brahma supercomputing facility at IISER Pune, which is part of the National Super Computing Mission under the Government of India; the authors are thankful for the support of the UChicago and IISER Pune computing teams.  This research made extensive use of \https{arXiv.org} and NASA's Astrophysics Data System for bibliographic information. YZ acknowledges the Aspen Center for Physics for its hospitality during the completion of this study, which is supported by the National Science Foundation under Grant PHY-1607611. YZ is supported by the Kavli Institute for Cosmological Physics at the University of Chicago through an endowment from the Kavli Foundation and its founder Fred Kavli.

\end{acknowledgments}

\appendix

\section{Estimator for $\Delta \Sigma(R)$ in Data}
\label{ref:appendix_data}

In this section we describe the estimator for $\Delta\Sigma(R)$ that is used to measure the weak lensing of galaxies. The estimator is given by,
\begin{equation}
    \Delta\tilde{\Sigma} (R) = \frac{\sum_{ij} s^{ij} \boldsymbol{{\rm e}}_{\rm t}^{ij} (R) }{\sum_{ij} s^{ij} \Sigma^{-1}_{\rm c,MC}(z^i_{\rm l},z^j_{\rm s}) \mathcal{R}^j} ,
\end{equation}
where $i$ corresponds to the lenses, $j$ the sources, $\mathcal{R}$ the shear response from \textsc{Metacalibration}, and 
\begin{equation}
    s^{ij} = \omega^j \Sigma^{-1}_{\rm c,mean}(z^i_{\rm l},z^j_{\rm s})
\end{equation}
is the weight for optimizing the measurement, where $\omega^j$ is the square inverse of the shear measurement error of the $j$-th source. 
Here, $\Sigma^{-1}_{\rm c,MC}$ denotes the inverse of critical density with the redshift of the source galaxy arbitrarily chosen from the probability distribution determined by the BPZ algorithm, and $\Sigma^{-1}_{\rm c,mean}$ that evaluated at the mean redshift.
In addition, to reduce the contamination from the foreground and cluster member galaxies, we select source galaxies whose photo-$z$'s lie above $\Delta z=0.1$ of the lens cluster.  

However, the measured weak--lensing profile around clusters still suffers from such contamination from the cluster member galaxies, for which one must correct the data. We estimate this systematic, called "boost factor" ($\mathcal{B}$), using $P(z)$ decomposition method \citep{Varga19}. In $P(z)$ decomposition method, one calculates the fraction of contamination in our weak--lensing estimator by decomposing the probability distribution of the source redshift into the true source distribution and the contamination part:
\begin{equation}
    P(z|R) = f_{\rm cl}(R) \, P_{\rm cont}(z|R) + (1-f_{\rm cl}(R)) \, P_{\rm bg}(z),
\end{equation}

where $f_{\rm cl}(R)$ represents the fraction of the contamination as a function of radius, $P_{\rm cont}(z|R)$ the redshift probability distribution of the cluster member galaxies at radius $R$ which is assumed to be Gaussian, $P_{\rm bg}(z)$ the probability distribution of the true background source redshift.
We calculate $P_{\rm bg}(z)$ around random points and $P_{\rm cont}(z|R)$ is assumed to be Gaussian. 
Then, given $P(z|R)$ and $P_{\rm bg}(z)$ from the data, the redshift distribution could be fit with $f_{\rm cl}(R)$ and the width of $P_{\rm cont}(z|R)$ being the free parameters.
The boost factor is related to $f_{\rm cl}$ as,
\begin{equation}
    \mathcal{B}(R) = \frac{1}{1-f_{\rm cl}(R)},
\end{equation}\\

and the corrected weak--lensing profile becomes\\
\begin{equation}
    \Delta\tilde{\Sigma}_{\rm corr} = \mathcal{B}\Delta\tilde{\Sigma}.
\end{equation}

The covariance matrix for the measurements is calculated with the jackknife resampling method, with 100 approximately equal-area patches. 

We refer readers to \cite{Varga19,McClintock19} for details of these estimators and validations thereof. Also, the details of the measurement can be found in \cite{DES:2021qzb}. Here we use the measurements to compare with the measurement of $\Delta \Sigma$ for different dark matter models.
\bibliographystyle{yahapj}
\bibliography{references,references1}

\end{document}